ESA Voyage 2050 White Paper

# GAUSS

**G**enesis of **A**steroids and Evol**U**tion of the **S**olar **S**ystem

A Sample Return Mission to Ceres

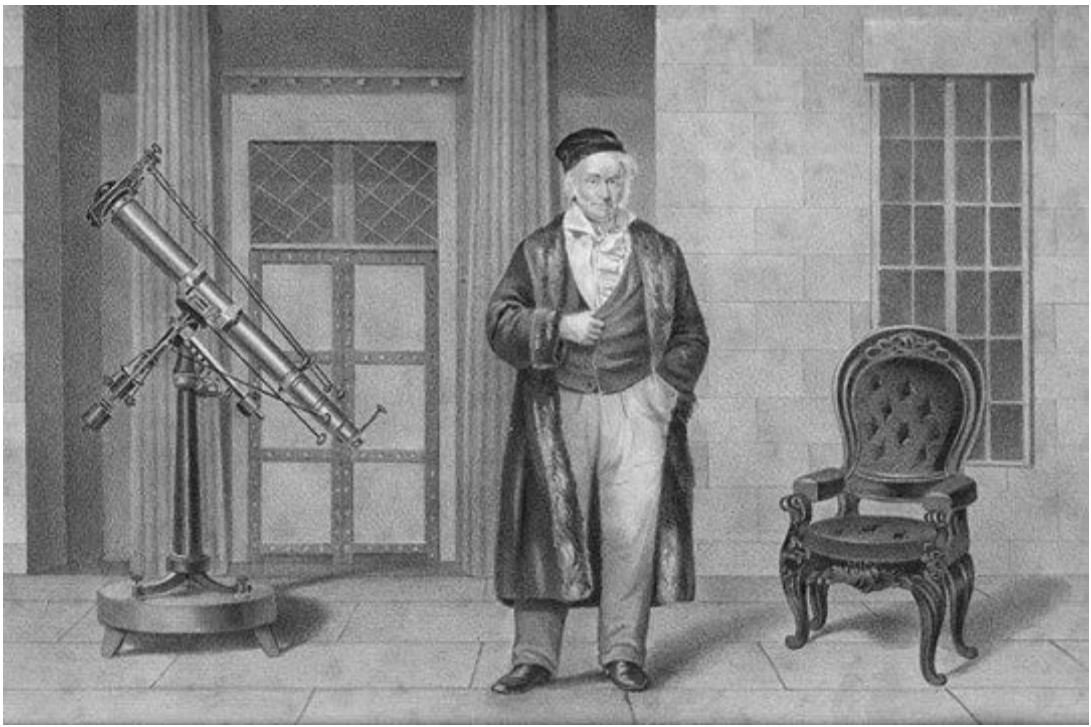

contact scientist

**Xian Shi**

Max Planck Institute for Solar System Research
Justus-von-Liebig-Weg 3, 37077 Göttingen, Germany

shi@mps.mpg.de

## Members of the core proposing team

| | |
|---|---|
| Julie Castillo-Rogez | NASA Jet Propulsion Laboratory, 4800 Oak Grove Dr, Pasadena, CA 91109, USA |
| Henry Hsieh | Planetary Science Institute, Tucson, AZ 85719, USA |
| Hejiu Hui | School of Earth Sciences and Engineering, Nanjing University, 163 Xianlin Avenue, Nanjing 210023, China |
| Wing-Huen Ip | Space Science Institute, Macau University of Science and Technology, Av. Wai Long, Macau |
| Hanlun Lei | School of Astronomy and Space Science, Nanjing University, 163 Xianlin Avenue, Nanjing 210023, China |
| Jian-Yang Li | Planetary Science Institute, Tucson, AZ 85719, USA |
| Federico Tosi | INAF-IAPS, Istituto di Astrofisica e Planetologia Spaziali, Via del Fosso del Cavaliere, 100, 00133 Rome, Italy |
| Liyong Zhou | School of Astronomy and Space Science, Nanjing University, 163 Xianlin Avenue, Nanjing 210023, China |
| Jessica Agarwal | Max Planck Institute for Solar System Research, Justus-von-Liebig-Weg 3, 37077 Göttingen, Germany |
| Antonella Barucci | LESIA, Observatoire de Paris, CNRS, UPMC Univ. Paris 06, Univ. Paris-Diderot, 5 Place J. Janssen, 92195 Meudon Principal Cedex, France |
| Pierre Beck | Univ. Grenoble Alpes, CNRS - Centre National de la Recherche Scientifique, IPAG, Grenoble, France |
| Adriano Campo Bagatin | Departamento de Física, Ingeniería de Sistemas y Teoría de la Señal. P.O. Box 99, 03080 Alicante, Spain |
| Fabrizio Capaccioni | INAF-IAPS, Istituto di Astrofisica e Planetologia Spaziali, Via del Fosso del Cavaliere, 100, 00133 Rome, Italy |
| Andrew Coates | Mullard Space Science Laboratory, University College London, Holmbury St Mary, Dorking, Surrey RH5 6NT, UK |
| Gabriele Cremonese | INAF - Osservatorio Astronomico di Padova, Vicolo dell'Osservatorio 5, 35122 Padova, Italy |
| Rene Duffard | Instituto de Astrofísica de Andalucía (CSIC), Glorieta de la Astronomía S/N, 18008-Granada, Spain |
| Manuel Grande | University of Aberystwyth, Penglais, Aberystwyth SY23 3FL, United Kingdom |
| Ralf Jaumann | DLR Institute of Planetary Research, Rutherfordstraße 2, 12489 Berlin, Germany |
| Geraint Jones | Mullard Space Science Laboratory, University College London, Holmbury St Mary, Dorking, Surrey RH5 6NT, UK |
| Esa Kallio | School of Electrical Engineering, Aalto University, Maarinkatu 8, P.O. Box 15500, FI-00760 Aalto, Finland |
| Yangting Lin | Institute of Geology and Geophysics, Chinese Academy of Sciences, 19 Beitucheng W Rd, Chaoyang Qu, 100029, Beijing, China |
| Olivier Mousis | Aix Marseille Université, CNRS, LAM (Laboratoire d'Astrophysique de Marseille) UMR 7326, F-13388 Marseille |
| Andreas Nathues | Max Planck Institute for Solar System Research, Justus-von-Liebig-Weg 3, 37077 Göttingen, Germany |
| Jürgen Oberst | DLR Institute of Planetary Research, Rutherfordstraße 2, 12489 Berlin, Germany |
| Adam Showman | Lunar and Planetary Laboratory, University of Arizona, 1629 E University Blvd, Tucson, AZ 85721, USA |
| Holger Sierks | Max Planck Institute for Solar System Research, Justus-von-Liebig-Weg 3, 37077 Göttingen, Germany |
| Stephan Ulamec | German Aerospace Center, DLR, 51147 Cologne, Germany |
| Mingyuan Wang | National Astronomical Observatory, Chinese Academy of Science, 20A Datun Road, Chaoyang District, Beijing, 1000012, China |

\* A list of members of the extended proposing team to be found at the end of the proposal.

# Executive summary

The goal of Project GAUSS is to return samples from the dwarf planet Ceres. Ceres is the most accessible ocean world candidate and the largest reservoir of water in the inner solar system. It shows active cryovolcanism and hydrothermal activities in recent history that resulted in minerals not found in any other planets to date except for Earth's upper crust. The possible occurrence of recent subsurface ocean on Ceres and the complex geochemistry suggest possible past habitability and even the potential for ongoing habitability. Aiming to answer a broad spectrum of questions about the origin and evolution of Ceres and its potential habitability, GAUSS will return samples from this possible ocean world for the first time. The project will address the following top-level scientific questions:

- What is the origin of Ceres and the origin and transfer of water and other volatiles in the inner solar system?
- What are the physical properties and internal structure of Ceres? What do they tell us about the evolutionary and aqueous alteration history of icy dwarf planets?
- What are the astrobiological implications of Ceres? Was it habitable in the past and is it still today?
- What are the mineralogical connections between Ceres and our current collections of primitive meteorites?

GAUSS will first perform a high-resolution global remote sensing investigation, characterizing the geophysical and geochemical properties of Ceres. Candidate sampling sites will then be identified, and observation campaigns will be run for an in-depth assessment of the candidate sites. Once the sampling site is selected, a lander will be deployed on the surface to collect samples and return them to Earth in cryogenic conditions that preserves the volatile and organic composition as well as the original physical status as much as possible.

# 1. Background

Though the Rosetta mission to comet 67P/Churyumov-Gerasimenko came to an end only three years ago in 2016, it is important to recall that the planning activity eventually leading to its approval by ESA was initiated more than three decades ago in 1983. An equally, if not more, ambitious project in the framework of "Voyage 2050" is proposed here. The target is the innermost dwarf planet, Ceres, which was discovered on New Year's Day of 1801, by the Italian astronomer Giuseppe Piazzi in Palermo Observatory. At the time of its discovery, Ceres was considered to be the missing planet between the orbits of Mars and Jupiter as predicted by the Titius-Bode law. Ceres' location was confirmed in December the same year using the orbital elements calculated by then 24-year-old Carl Fredrich Gauss (Teets & Whitehead, 1999). The name of the project with the acronym of GAUSS for "the Genesis of Asteroids and the Evolution of the Solar System" is partly a tribute to this scientific episode of great importance in astronomy and planetary science.

# 2. Scientific Rationale

## 2.1 Ceres in the history of the Solar System

The oldest solids found in the solar system have an age of 4567.5 Myr (Connelly et al., 2008). After that, the formation of gas giants must finish before the dispersal of the protoplanetary gas disk, which lasts 2-10 Myr. Then, about 30-100 Myr later, the terrestrial planets formed in the planetesimal disk (Jacobson et al., 2014).

The planetesimal disk not only provides all the materials that constitute the terrestrial planets (and also the core of gas giants), but also exerts perturbation force on the growing and grown planets. This perturbation force may continuously modify the planetary orbits in a relatively gentle way (Fernandez & Ip 1984). Before the dispersal of the gas disk, planets may also experience much quicker migration due to the interaction with the massive gas disk. During the quick planetary migration, some violent mechanisms, e.g. low-order mean motion resonance between planets (Tsiganis et al., 2005; Walsh et al., 2011), bring up abrupt variations to the orbital configuration of planets. The orbits of planets might expand or shrink by large distance, the planets were rudely thrown into the outer planetesimal disk, and planets or planet embryos may be ejected to extremely faraway orbits or out of the solar system.

In these either mild or wild processes, the planetesimal disk was stirred up, and materials in the disk may migrate along with the planets, or transport by large distance via chaotic routes in between planetary orbits. Planetesimals may accrete or fragment by collisions. Planet embryos arose and crashed, producing the great diversity of celestial bodies. The interaction between the planetesimal disk and planets/protoplanets is not fully understood yet. The growth and evolution of terrestrial planets and asteroids (in a wide radial range from Mars orbit to Kuiper belt) in the early stages of the solar system are still of great uncertainty. Various scenarios have been proposed to reproduce this early history.

Given that their properties depend on their formation circumstances and evolutionary processes experienced since their formation, small bodies (e.g., asteroids, comets, and trans-Neptunian objects) in the solar system are of interest to researchers because they provide a way to probe the protoplanetary disk from which our solar system formed (e.g., Scott et al., 1989; Jones et al. 1990; Bottke et al., 2002). By determining the chemical and physical properties of various small bodies in our solar system, we can gain insight into chemical and thermal conditions in different areas of the protoplanetary disk and also investigate the chemical, thermal, collisional, and dynamical processes that have shaped those populations since their formation. A successful scenario of planet migration should provide the mechanisms that were able to efficiently deliver planetesimals from different zones to the main belt. In addition, the structure of the main belt, the mass depletion, the proper excitation of orbital eccentricities and inclinations, should all be reproduced.

One of the earliest attempts to map the compositional structure of our asteroid belt and use it to infer information about the origin and evolution of the solar system was performed by Gradie & Tedesco (1982). They found a systematic distribution of compositional types of asteroids in the main asteroid belt that they suggested was consistent with chemical condensation models of the solar system (Fig. 1). The authors concluded that it was unlikely that this distribution could be explained by the chaotic transport of objects from different regions of the solar system into the asteroid belt, and proposed instead that it indicated that the asteroids formed at or near their present locations.

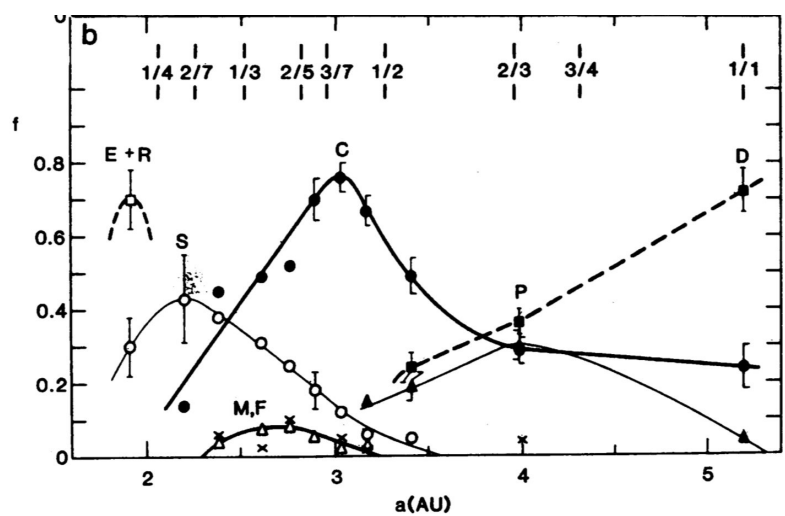

**Figure 1.** Observed relative numerical distribution of taxonomic types for a bias-corrected sample of 656 main-belt asteroids where smooth curves have been drawn through the data points of each taxonomic type to more clearly show their relative distributions (From Gradie & Tedesco, 1982).

The work of Gradie & Tedesco (1982) has since undergone a major update in the form of a study by DeMeo & Carry (2013), who used Sloan Digital Sky Survey (SDSS) Moving Object Catalog data to derive taxonomic classifications for about 35 000 objects, including objects as small as 5 km in diameter. The inclusion of much smaller objects in this sample than were previously available as well as the computation of compositional distributions by mass (instead of by number; Fig. 2) led DeMeo & Carry (2014) to conclude that the asteroid belt preserves a history of solar system evolution that is far more complex than previously thought (Fig. 3), with asteroids of various taxonomic types scattered throughout the asteroid belt, including in regions where they are not expected based on dynamically static solar system formation models. Instead, the history of the solar system as recorded by main belt asteroids likely includes relatively brief periods of dramatic mixing caused by giant planetary migration (e.g., those proposed as parts of the well-known Nice and Grand Tack models; Tsiganis et al., 2005; Morbidelli et al., 2005; Gomes et al., 2005; Levison et al., 2009, 2011; Walsh et al., 2011) as well as less dramatic but ongoing processes such as collisions and small body migration due to mean-motion resonances with the giant planets and the Yarkovsky effect (e.g., Bottke et al., 2005, 2006; Gladman et al., 1997; Farinella et al., 1998).

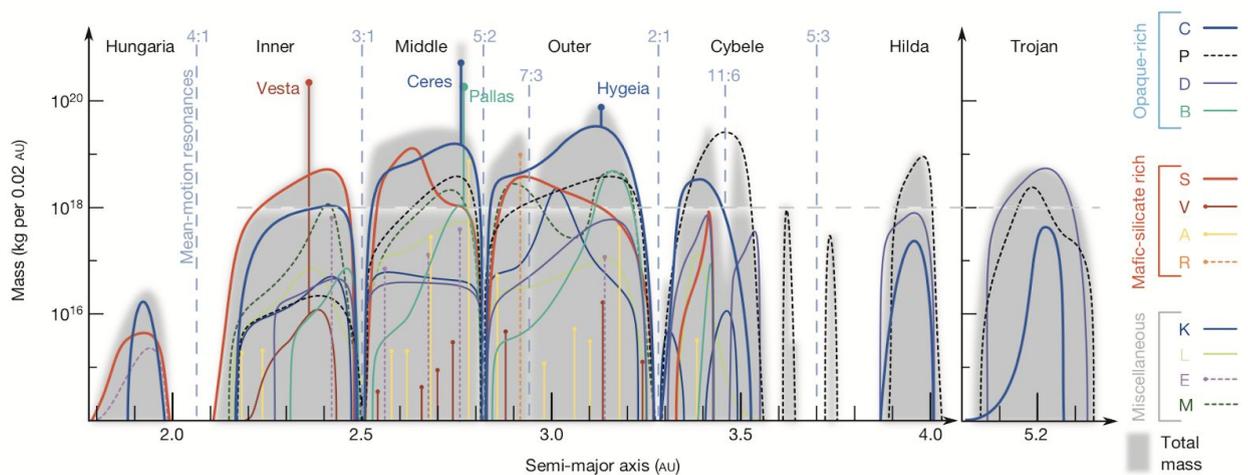

**Figure 2**. The compositional mass distribution of small bodies from the main asteroid belt through the Jovian Trojan population, where the gray background indicates the total mass of small bodies in each semimajor axis bin and each color represents a certain spectral class of asteroid, as labeled (from DeMeo & Carry, 2014).

The fraction and composition of ice in a body is of particular interest due to the strong temperature constraints they provide and the importance of water to life on Earth. The so-called "snow line" refers to the distance from the Sun at which the temperature dips below the condensation temperature of water, causing it to freeze into solid ice and then become incorporated into accreting planetesimals. The exact location of the snow line in our protoplanetary disk was dependent on a variety of poorly constrained environmental conditions including opacity, mass density, and accretion rate in the disk, and is also thought to have shifted with time as planetesimal accretion progressed and the aforementioned properties of the disk changed.

Bodies formed in the outer solar system such as between Jupiter and Neptune (the original accretion zone of current Oort Cloud objects; Hahn & Malhotra 1999; Weissman 1999) and beyond the orbit of Neptune (the Kuiper Belt) are well beyond the snow line and thus are certainly icy. Closer to the Sun, the situation is less certain. Observations of asteroids suggest that the snow line probably existed around 2.5 au from the Sun (Gradie & Tedesco 1982; Jones et al. 1990), but theoretical studies (e.g., Sasselov & Lecar 2000; Ciesla & Cuzzi 2006; Lecar et al. 2006) have placed it as close in as the orbit of Mars, or even closer. If this is true, it means that objects throughout the main asteroid belt could have incorporated water ice at the time of their formation.

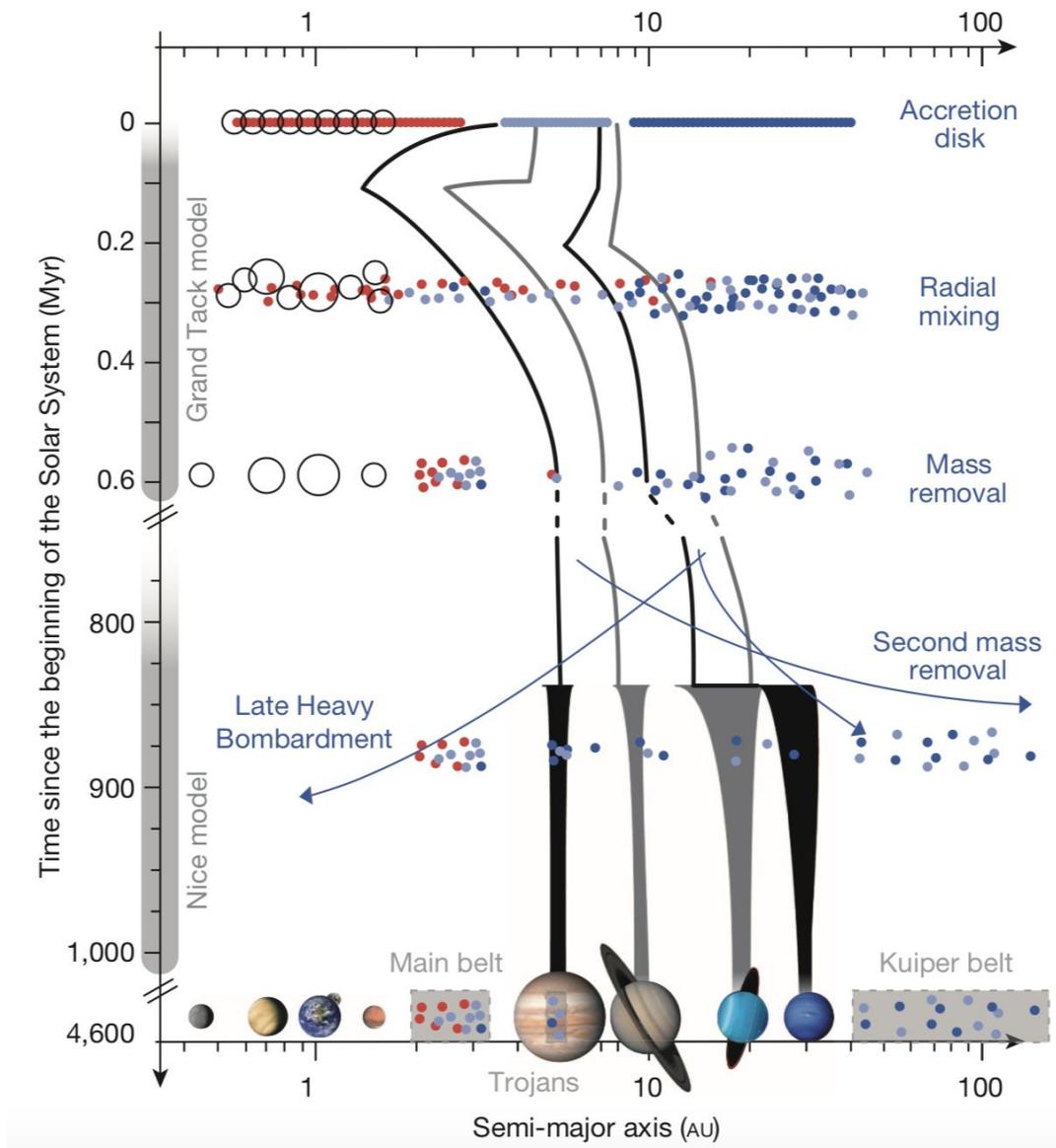

**Figure 3**. An illustration of the potential dynamical history of the solar system and the effects on its small body population based on proposed planetary migration models (From DeMeo & Carry, 2014).

Evidence of past and present-day ice has in fact been found in main-belt asteroids. Studies of meteorites linked to the asteroid belt as well as remote spectroscopic observations of asteroids have revealed the presence of aqueously altered minerals, indications that liquid water was once present (e.g., Hiroi et al. 1996; Keil et al. 2000, Rivkin et al. 2002). Meanwhile, spectroscopic evidence of water ice frost has been reported for various main-belt asteroids including (24) Themis (e.g., Rivkin & Emery 2010; Takir et al. 2012; Hargrove, et al. 2015), while some main-belt objects have even been observed to exhibit comet-like activity that has been attributed to the sublimation of present-day volatile ices, i.e., the so-called main-belt comets (Hsieh & Jewitt, 2006). The location, abundance, distribution, and inferred water content of currently and formerly icy main-belt objects provide valuable clues for discerning the primordial location and evolution of the snow line, although this of course is also complicated by the aforementioned likely transport of at least some small bodies from their original formation locations due to giant planet migration and other dynamical processes.

The bulk-rock isotope anomalies of meteorites reveal an isotopic dichotomy: two distinct trends defined by the non-carbonaceous (NC) bodies and the carbonaceous (CC) bodies respectively (Fig. 4). The CC bodies include carbonaceous chondrites, a few ungrouped achondrites, and IIC, IID, IIF, IIIF, IVB iron meteorites. The NC bodies include Earth, the Moon, Mars, ordinary chondrites, enstatite chondrites, and most of achondrites. This dichotomy has been further observed in Mo, W, Ru, and Ni isotopic systematics (Budde et al., 2016; Kruijer et al., 2017; Nanne et al., 2019; Poole et al., 2017). However, the origins of the isotopic anomalies and their correlations are unknown, although 54Cr-enrichment can be attributed to nano-sized Cr-oxide probably ejected from supernova (e.g. Qin et al., 2011). Warren (2011) proposed that the CC bodies formed in the outer solar system and the NC bodies formed in the inner solar system. The formation of Jupiter blocked the material exchange between the two groups, resulting in the isotopic dichotomy (Warren, 2011). As a consequence of radial migration and mass growth of the giant planets in the solar system, some CC planetesimals and embryos moved inwards from the outer solar system to the main belt (Walsh et al., 2011; Warren, 2011). These CC materials eventually formed the parent bodies of carbonaceous chondrites.

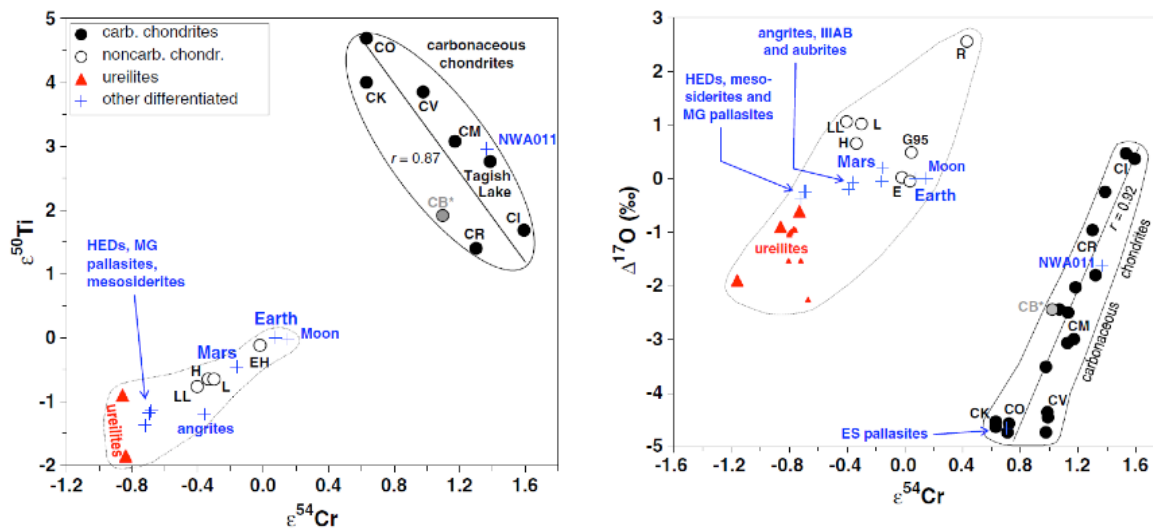

**Figure 4**. Dichotomy of stable isotopes among planetary materials (from Warren, 2011).

Ceres, the largest object in the main belt, preserves many clues to the formation and evolution of the main belt, as well as of the planetary system.

The surface material of Ceres is similar to carbonaceous chondrite though no meteorite in our collection has ever been linked with Ceres (Rivkin et al., 2011). Ceres may be a member of CC-group or has accreted in the margin region between CC and NC groups. Its large size probably kept Ceres from being merged by Jupiter and hence survived. Based on the data from the Dawn mission, recent geochemical simulations show that the surface mineralogy of Ceres is consistent with the aqueous alteration of CM chondrite (Neveu and Desch, 2015). Therefore, Ceres could be a carbonaceous body, isotopically different from the NC bodies that formed in the inner solar system (Warren, 2011). This suggests that Ceres migrated to the current neighborhood from the outer solar system. This origin is consistent with abundant ammonia detected on Ceres by the Dawn mission.

The origin of Ceres from the outer solar system suggested by remote sensing observation still requires decisive evidence for confirmation. The stable isotope anomalies of samples from Ceres could provide ground truth on where it formed and how it traveled (Fig. 4). Furthermore, all the geochemical information of samples could be used to understand the evolution of Ceres.

## 2.2 Results from the Dawn mission

Dawn mission performed a 3.5 years rendezvous with Ceres starting from early 2015, mapping its global topography, geomorphology at resolutions down to 35 m, surface mineralogy at resolutions down to 90 m, the top layer elemental abundance at effective resolutions of tens of km, as well as the gravity field to 18 degrees of spherical harmonics. During the last mission phase, Dawn entered a long elliptical orbit with an altitude of 35 km at periapsis for a detailed study of the narrow longitudinal strip at about 240º E that stretches from Occator crater to Urvana crater.

Dawn results confirmed that Ceres is volatile-rich, has a partially differentiated interior, and has experienced global aqueous alteration. Dawn revealed Ceres as a geologically active dwarf planet with cryovolcanism and geothermal activity in the recent history of a few million years and probably even at the present.

Ammoniated phyllosilicates are distributed all over the surface of Ceres (De Sanctis et al. 2015, Ammannito et al. 2016). The ubiquitous distribution of hydrated minerals suggests widespread aqueous alteration on Ceres. The incorporation of ammonia in Ceres' surface mineralogy is an indication that Ceres must have accreted materials from the giant planet formation region and/or outer solar system. However, it is not clear whether Ceres is formed in the outer solar system (McKinnon et al. 2012), or formed in situ but accreted materials radially transported from the outer solar system.

The top surface layer of Ceres is rich in water of hydration and water ice, and the abundance of water ice is relatively low towards the equator and high towards the polar regions (Prettyman et al. 2017). Such a latitudinal variation is consistent with the evolution of subsurface water ice as controlled by the thermal condition on the surface and shallow subsurface of Ceres (Schorghofer 2008, 2016) and the low obliquity of Ceres (Russell et al. 2016). Water ice has also been directly identified in about ten specific places at the surface, at latitudes poleward of 30° in fresh craters near rim shadows (Combe et al. 2016, 2019) and inside the permanently shadowed craters in the polar regions (Platz et al. 2016, Schorghofer et al. 2016).

The abundant water ice on Ceres (Fig. 5) leads to differentiation and aqueous alteration that shaped the mineralogical composition of Ceres in its crust and mantle. The presence of carbonates and ammonium salts are previously found only on Earth and Enceladus (De Sanctis et al. 2016). The existence of hydrated salts played a key role in driving the geological process on Ceres, including cryovolcanism (Ruesch et al. 2016, 2019) and geothermal activity (Scully et al. 2019b) in its geologically recent history.

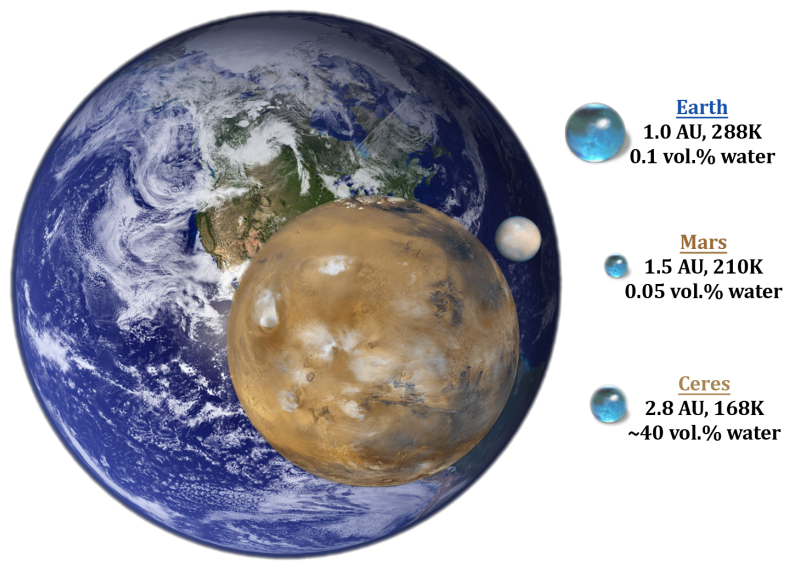

**Figure 5.** Comparison view of water abundances on Earth, Mars, and Ceres.

The most prominent geomorphological feature that is considered of cryovolcanic origin is Ahuna Mons (Fig. 6) (Ruesch et al. 2016, 2019). Its distinct size, shape, and morphology are consistent with being a volcanic dome formed by extrusions of highly viscous melt-bearing material. At the summit is the buildup of a brittle carapace, which partially fractured and disintegrated to generate the slope debris. The bright streaks in the slope debris are rich in Na-carbonates (Zambon et al., 2017). The gravitational relaxation of the enclosed ductile core shapes the overall topographic profile of the summit, requiring an extruded material of high viscosity. The age of the most recent activity on Ahuna Mons is about 210 ± 30 million years. Other possible cryogenic features include smaller domes (Sori et al. 2017, 2018), fractures in the crater floor (Buczkowski et al., 2016), post-impact modification by the deposition of extended plains material with pits and widely dispersed deposits that form a diffuse veneer on the preexisting surface (Krohn et al., 2016), and multiple lobate flows (Schmidt et al., 2017).

Shallow subsurface volatiles are also evident from the many other geomorphological features, such as the pitted terrain (Sizemore et al., 2017). On the other hand, the crater morphology and the simple-to-complex crater transition indicate that Ceres' outer shell is likely neither pure ice nor pure rock, but a mixture of ice, rock, salts and/or clathrates that allows for limited viscous relaxation that varies spatially (Hiesinger et al., 2016; Bland et al., 2016).

The distinctive bright regions within Occator crater are one of the most remarkable features on Ceres observed by Dawn (Fig. 7). The Occator crater is about 90 km in diameter, hosting the bright deposit covering the pit-dome complex named Cerealia Facula in the center, and a group of secondary bright deposit named Vinalia Faculae on the east side of the crater floor. While Ceres' average surface contains Mg- and Ca-carbonates and Mg- and $NH_4$-phyllosilicates, the Occator's faculae contain Na-carbonate, Al-phyllosilicates, and $NH_4$-chloride (Raponi et al., 2019). Theoretical modeling suggested the possibility of a brine reservoir beneath Occator crater, and the gradual freezing of this reservoir was the driver of briny cryolavas (Quick et al., 2019). The existence of $NH_4$-salts essentially lowered the eutectic point of brine and extended the duration of cryovolcanic activity on Ceres. Laboratory studies on the freezing process of ammonium-sodium-carbonate-chloride brines showed that slow-freezing (<30 K/min) is most compatible with the observed composition of brines observed in Occator crater (Thomas et al., 2019). A variety of morphological features are observed in the crater. The presence of linear and concentric fractures on the crater floor is associated with cryomagmatic intrusion, and degassing or desiccation processes for the volatile rich Occator ejecta (Buczkowski et al. 2019). The cross-cutting relationship between stratigraphic units indicate that the Cerealia Facula were emplaced prior to the formation of Occator's central pit and dome, and multiple episodes of emplacement could have happened (Scully et al., 2019). Crater counting in the Occator ejecta results in an age of about 20 million years (Neesemann et al., 2019), and the age of faculae are a few millions of years (Nathues et al. 2017, 2019).

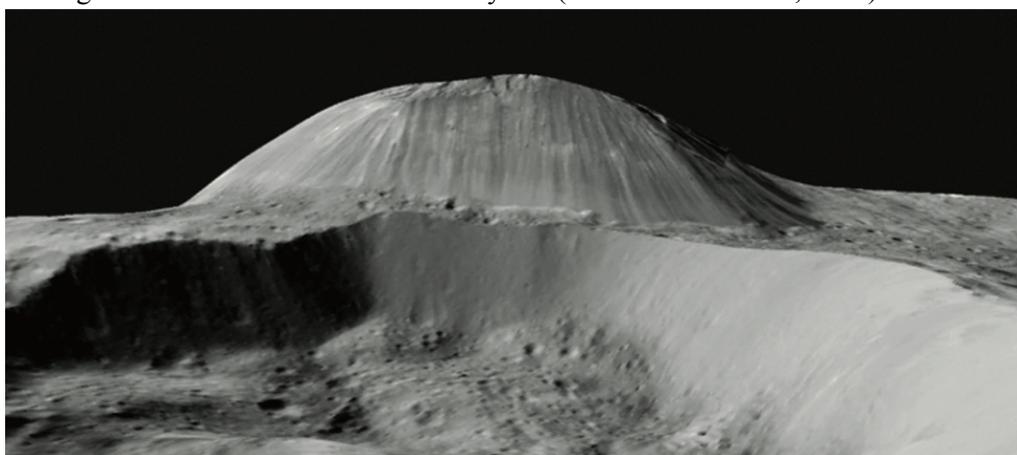

**Figure 6**. Perspective view of Ahuna Mons on Ceres from Dawn Framing Camera data (no vertical exaggeration). The mountain is 4 km high and 17 km wide in this south-looking view. (Reproduced from Ruesch et al. 2016).

Based on these results, a time sequence of the formation of the Occator bright deposit was postulated (Scully et al., 2019b). The outer, diffuse edge of Cerealia Facula was emplaced shortly after the Occator-forming impact, either by impact-induced hydrothermal brine deposition or by salt-rich water fountaining, perhaps sourced in a pre-existing reservoir. The majority of the Cerealia Facula formed from the evaporation of brine reaching the surface after the collapse of central pit. Cryomagmetic intrusion uplifted the central dome and the associated fracture systems. Finally, brine continued to ascend to the surface through fractures and form the Vinalia Faculae. The bright deposits on Ceres is perhaps the most direct evidence of geologically recent cryovolcanic and geothermal activity on this dwarf planet.

The presence of water ice deposits on the surface of Ceres, as well as the widespread distribution of shallow subsurface water has been associated with the active outgassing of Ceres previously observed from Earth orbit or the Earth-Sun L2 Lagrange point(A'Hearn & Feldman 1992; Küppers et al. 2012). However, the observed surface and subsurface water ice on Ceres do not appear to be sufficient to supply the observed water production rate (Schorghofer et al. 2017; Landis et al. 2017, 2019). Multiple attempts to detect water outgassing around Ceres also failed, suggesting that water outgassing from Ceres is variable but also does not appear to depend on heliocentric distance, although hypothesis has been proposed that the solar energetic particles might be a cause (Villarreal et al., 2017). The water loss mechanisms on Ceres and the characteristics of its transient water exosphere are therefore still completely obscured.

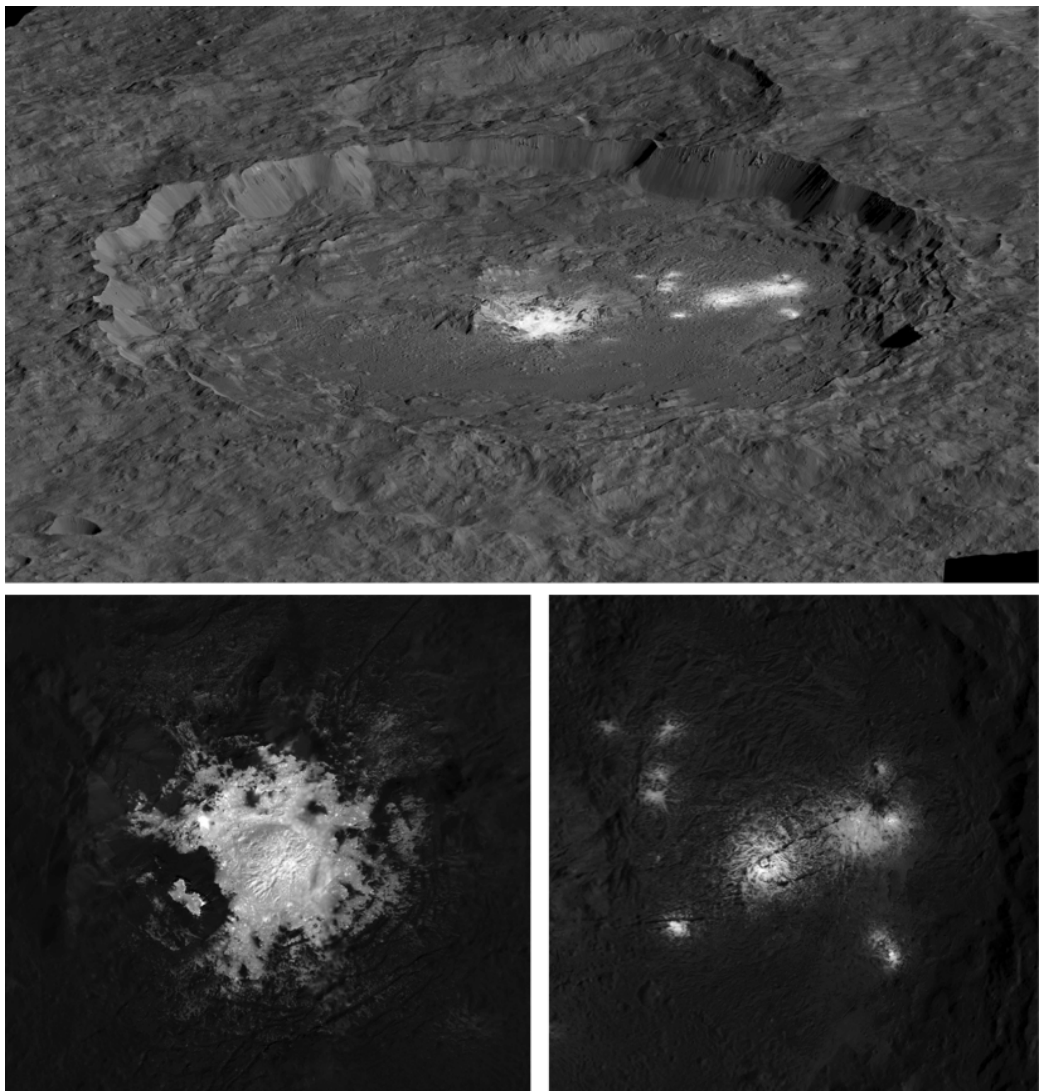

**Figure 7**. Top: Perspective view of Occator crater from the south and the bright deposits from Dawn Framing Camera data. The Cerealia Facula is saturated in this brightness stretch. Bottom left: Zoom-in view of the Cerealia Facula at the center of Occator crater shows the details of the bright deposit. Bottom right: Zoom-in view of the Vinalia Faculae.

Gravitational data and geophysical modeling suggested that Ceres is partially differentiated to a ~40 km thick crust composed of rock, ice, salts, and/or clathrates with no more than 30% water ice (Ermakov et al., 2017; Bland et al., 2016; Fu et al., 2017). The crust is relatively strong and limited the relaxation of small craters. Below the crustal layer is a denser rocky mantle with a relatively weak upper layer with brine-filled pore space that controls the global shape of Ceres (Fu et al., 2017). The possibility of a dehydrated core below 100 km cannot be ruled out (King et al., 2018).

Near-infrared data returned by the Visible and InfraRed mapping spectrometer (VIR) on board the Dawn spacecraft first detected an organic absorption feature at 3.4 micrometers on Ceres. This signature is diagnostic of organic matter and is mainly localized on a broad region of ~1000 square kilometers close to the ~50-kilometer Ernutet crater (De Sanctis et al., 2017). The shape of the 3.4-μm band and the lack of an associated 3.25-μm feature could exclude organics with a high content of aromatic carbon such as anthraxolites as main carriers of the features on Ceres, in favor of hydrocarbons rich in aliphatic carbon (like asphaltite and kerite) (De Sanctis et al., 2017). However, a firm identification of the specific organic material responsible for the 3.4-μm absorption band observed close to crater Ernutet is missing, and the assessment of its nature (endogenous vs. exogenic) is still debated, making this a compelling scientific case for any future space mission to Ceres. Furthermore, in principle organic compounds might exist also on top of Cerealia Facula, the brightest spot located roughly in the middle of crater Occator, even though they were not revealed by near infrared data acquired by VIR even at pixel resolution as high as about ten meters per pixel, which was achieved towards the end of the Dawn mission.

The cryovolcanism and geothermal activity in the recent history of Ceres, the existence of liquid water on a global scale, and the possibility of liquid brine pocket at the present suggest an active planet that could have strong astrobiological significance. The combined presence on Ceres of ammonia-bearing hydrated minerals, water ice, carbonates, salts, and organic material, revealed a very complex chemical environment, suggesting favorable environments to prebiotic chemistry in a subsurface aqueous environment (De Sanctis et al., 2017). In this respect, the ability to detect, determine, and quantify any organics on Ceres is a clear step toward assessing habitability.

Dawn's observations confirmed earlier predictions for a volatile-rich crust encompassing the bulk of a former ocean, now frozen, and provide hints for a weak interior that may reflect the presence of a relict liquid layer or brine pockets. These observations led to Ceres' classification as a "candidate" ocean world in the Roadmap for Ocean Worlds (Hendrix et al. 2019). Current knowledge indicates that Ceres once had water, organic building blocks for life, energy sources, and redox gradients, and perhaps still does today. Perhaps more importantly, Ceres' astrobiological value comes from its potential for continuous habitability, commencing directly after accretion with a global ocean in which advanced chemical differentiation developed. This global ocean could have been maintained for billions of years (Travis et al. 2018). Most of Ceres' surface properties record the consequences of that early period, while contemporary activity is evident in a few places. However, per its size and water abundance, Ceres belongs to a class of objects that could host a high fugacity of hydrogen, organic molecules, and alkaline conditions, as was suggested for Europa (e.g., McKinnon and Zolensky 2003) and inferred from Cassini observations of Enceladus (Postberg et al. 2011; Marion et al. 2012).

Future exploration of Ceres would reveal the degree to which liquid water and other environmental factors may have combined to make Ceres a habitable world. Confirmation of the existence of liquid inside Ceres at present, and assessment of its extent, is the natural next step when following the roadmap for ocean worlds. Another key question for any coming mission is whether there exist oxidants at the surface to help the emergence of life or at least its habitability. A favorable redox discovery would deepen the case for Ceres meeting our current definition of habitability. Another topic of major importance is about quantifying the abundance, sources and sinks, and chemical forms of CHNOPS elements and assessing the inventory and determining the origin(s) of organic compounds beyond the limited observations of Dawn and determining their origin(s).

In summary, as illustrated in Fig. 8, Dawn revealed great scientific significance of Ceres:
- Rich in water ice and other volatiles relevant to our understanding of the history of water ice and volatiles in the inner solar system
- Formation and evolutionary history representative to ice-rich objects in the outer solar system
- Geologically active with minerals present only on Earth and Enceladus
- Closest and most accessible planet with cryovolcanism and geothermal activity
- Ammonia-bearing hydrated minerals, water ice, carbonates, salts, and organic matter make a complex chemical environment that could favor prebiotic chemistry

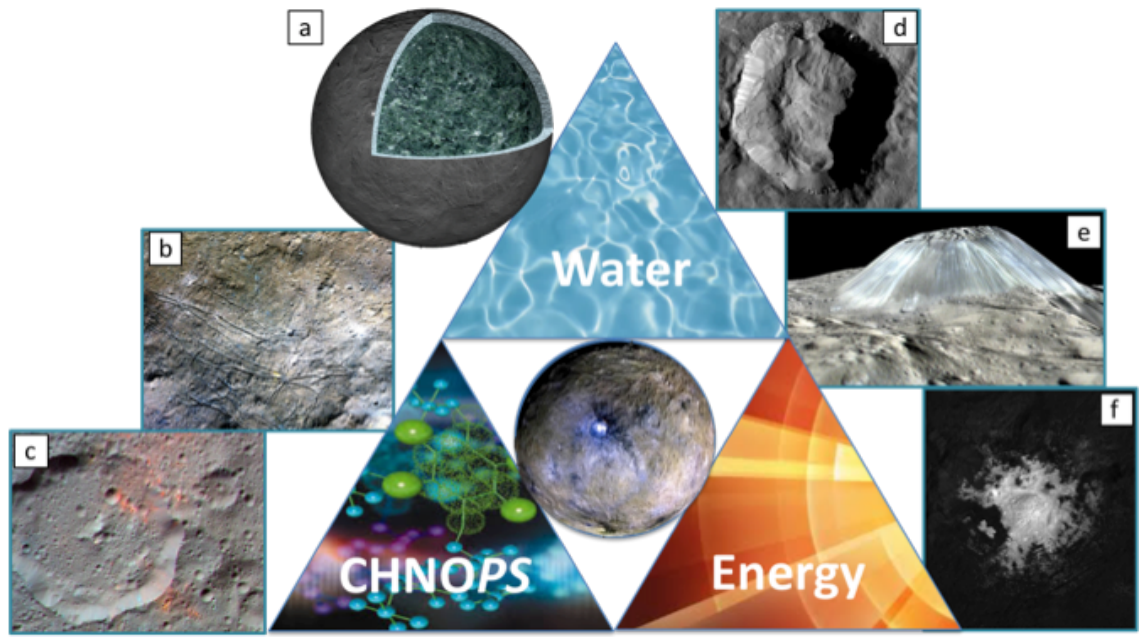

**Figure 8.** Summary of Dawn's observations of Ceres addressed in the text. [Credit for individual images: NASA/JPL/Caltech/IAPS/MPS/DLR/INAF/ASI.
(a) Geophysical data confirmed the abundance of water ice and the need for gas and salt hydrates to explain the observed topography and crustal density.
(b) Various types of carbonates and ammonium chloride have been found in many sites across Ceres' surface (e.g., salts exposed on the floor of Dantu crater)
(c) Ernuter crater (~52 km, above) and its area present carbon species in three forms (reduced in CxHy form, oxidized in the form of carbonates and intermediate as graphitic compounds.)
(d) Ceres shows extensive evidence for water ice in the form of ground ice and exposure via mass wasting and impacts (Left: Juling crater, ~20 km)
(e) Recent expressions of volcanism point to the combined role of radiogenic heating and low-eutectic brines in preserving melt and driving activity (Left: Ahuna Mons, ~4.5 km tall, ~20 km diameter)
(f) Impacts could create local chemical energy gradients in transient melt reservoirs throughout Ceres' history (Left: Cerealia Facula, ~14 km diameter)

## 2.3 Need for Ceres sample return

Vesta has now joined Mars and the Moon as the best understood extraterrestrial bodies, due at least in part to the fact that we have samples of all of them to study in the laboratory. Ceres can join that group with sample return.

Some exploration and scientific arguments for Ceres sample return are:
- Although carbonaceous chondrites provide the best analog for Ceres (McSween et al., 2018), we have no Ceres meteorites. The value of having samples for proper calibration of flight instruments and rigorous interpretation of remote sensing data is illustrated by studies of Vesta, Mars, and the Moon.
- Global spectral mapping of Ceres by Dawn's VIR demonstrates that its surface is covered almost everywhere by the same assemblage (ammoniated clay, serpentine, carbonate, and a darkening agent), but in slightly differing proportions (Ammannito et al., 2016). Thus, a representative regolith sample can be collected from nearly any location on the surface.
- The ice table occurs below several meters depth near the equator, and approaches the surface at higher latitudes (Prettyman et al., 2017). Sampling the regolith in the equatorial region would thus not require cryogenic collection and return (which is probably not possible anyway based on current technology).
- Ceres is more heavily altered by aqueous processes than any carbonaceous chondrites, and is differentiated with separation of ice and rock (McSween et al., 2018). At one time, it must have been

- an ocean world (Castillo-Rogez et al., 2018). We have never before sampled an ocean world, and our understanding of its alteration is based only on models.
- Remote sensing, at best, can only identify a few coexisting minerals. Other phases in minor proportions in a returned sample could constrain the conditions of alteration. It is also likely that Ceres regolith may contain some amorphous phases that could not be characterized by remote sensing.
- There is presently some controversy about the extent that Ceres' surface has been contaminated by exogenic chondrite impactors. These could be recognized and quantified by petrologic examination of a return sample.
- Ceres' chronology is based on crater size distribution analysis, which is dependent on the hypothesis of impactor flux (e.g. Marchi et al., 2010), leaving the absolute ages uncertain. Radiometric dating of a sample from a mapped geologic unit (Williams et al., 2018) could help define Ceres' chronology and the impactor flux in the asteroid belt.
- At a few locations on Ceres, notably Ahuna Mons and Occator Crater, recent brines have erupted and deposited salts (sodium-carbonate, ammonium chloride, plus more phases that have not been identified (De Sanctis et al., 2016; Zambon et al., 2017). Sampling a cryovolcanic site could be challenging, but impacts may have distributed these materials more widely. Small quantities of these phases, which provide important constraints on the nature of subsurface fluids, may occur in regolith samples close to such outcrops.
- Organic matter discovered at one location on Ceres (De Sanctis et al., 2017) suggests that it should be widely distributed in lesser amounts, although in some areas it may have been converted to graphite through UV irradiation. Understanding the organic component on Ceres has important implications for prebiotic chemistry and astrobiology.
- We do not know whether Ceres formed near its present location in the asteroid belt, or formed in the giant planet region and was later perturbed into the main belt by giant planet migration. Measurement of isotopes of H, C and N will place constraints on the origin of water and organics. Measurement of stable isotopes of oxygen, chromium, titanium, etc. (Scott et al., 2018) can place Ceres into its proper formation setting, as it has for other bodies for which we have samples.
- Comets from the outer solar system have long been suspected as the source of Earth's water. However, while a recent deuterium-to-hydrogen (D/H) ratio measurement for a comet is compatible with terrestrial ocean water (Hartogh et al. 2011), most comet D/H measurements are not (e.g., Altwegg et al. 2015). Meanwhile, dynamical studies (cf. O'Brien et al. 2018) indicate that large quantities of Earth's water could have been supplied by objects from the region of the solar system coinciding with the present-day outer asteroid belt. If Ceres can be determined to have formed in situ, a D/H ratio (in addition to other isotopic ratios) measured for Ceres would be very valuable in helping to assess the plausibility of the main asteroid belt as a source of the terrestrial water that is so critical to the rise of life on Earth.

## 3. Mission scenarios

### 3.1 Overview

The mission concept came about during a round-table discussion at the 4th Lunar and Deep-Space Exploration International Conference between July 22 and 24, 2019, in Zhuhai, China. One important objective of this proposal is therefore to promote scientific cooperation of the Chinese planetary science community with its European counterpart, taking advantage of the momentum of CNSA and the spirit of the "Voyage 2050 Initiative" of ESA. It might begin with a joint assessment study co-sponsored by both agencies or their representatives. The assessment should identify the scientific objectives, technical requirements, mission architecture and possible division of responsibilities and authorities.

If we follow the long-term trend in the development of scientific missions to the Moon and Mars, respectively, and that of asteroidal exploration, we would probably reach the conclusion that they might likely converge on a large-scale international space program of Ceres. The Dawn mission of NASA has yielded many exciting results but the scientific observations were limited to just three remote-sensing instruments, namely, the imaging optical camera, the near infrared imaging spectrometer, and the gamma-ray spectrometer/neutron detector. A more comprehensive payload on the orbiter(s) is needed for a full characterization and understanding of the surface and atmospheric/exospheric environment of Ceres. In addition, just like in the case of lunar or Mars exploration, different types of platforms such as lander and rovers might be required to provide critical information. A sample return mission might be regarded as the final step. Furthermore, because of Ceres' water-rich and organic-rich composition, it could become an important base for the establishment of research stations in support of deep space exploration.

As implied by "Dawn", the name of the first mission to Ceres, and the name of the present mission proposal "GAUSS", the in-depth exploration of Ceres should be viewed in terms of detailed investigations of the genesis and evolution of the asteroid belt and the solar system. The contribution of future in-situ and sample return explorations of Ceres in the context of studying the solar system ocean worlds and their habitability is illustrated in Fig. 9. It can therefore be envisaged that besides orbiter observations, a set of lander(s) and rover(s) would be needed. It is also possible that this program can be composed of a series of missions over a period of time to address different specific scientific questions.

The level of ESA engagement could vary between an M-class mission to an L-class mission. For example, even though our main focus is to initiate the planning of an L-class sample return mission, it could be streamlined to a Rosetta/Philae lander style or an ExoMars/Rosalind Franklin rover style mission for cost reason and expediency. The cost envelope might hence be fitted within the budget of an M-class project depending on the mission component to be chosen. Similar consideration would probably be pursued by CNSA also if it agrees to support a joint assessment study. It is important to emphasize that a sample return mission might still be possible with the participation of additional national agencies, besides CNSA and ESA as postulated here. Possible mission scenarios are summarized in Tab. 1.

**Table 1.** Summary of mission scenarios

| Mission type | Mission component | Mission class |
|---|---|---|
| Sample return | Orbiter+lander+return capsule | L-class |
| | | M-class with significant contribution from CNSA |
| Landing/Roaming | Orbiter+lander/hopper/rover | M-class with possible lander/rover contribution by CNSA |

The scientific goals and related measurement objectives of Project GAUSS are summarized in the traceability matrix (Tab. 2). The four scientific goals are: the origin of Ceres, its evolution and current status, its habitability, and its connection to the carbonaceous meteorite collections. The results from these investigations will have direct implications on our understanding of the evolution of the solar system and in particular of the icy satellites.

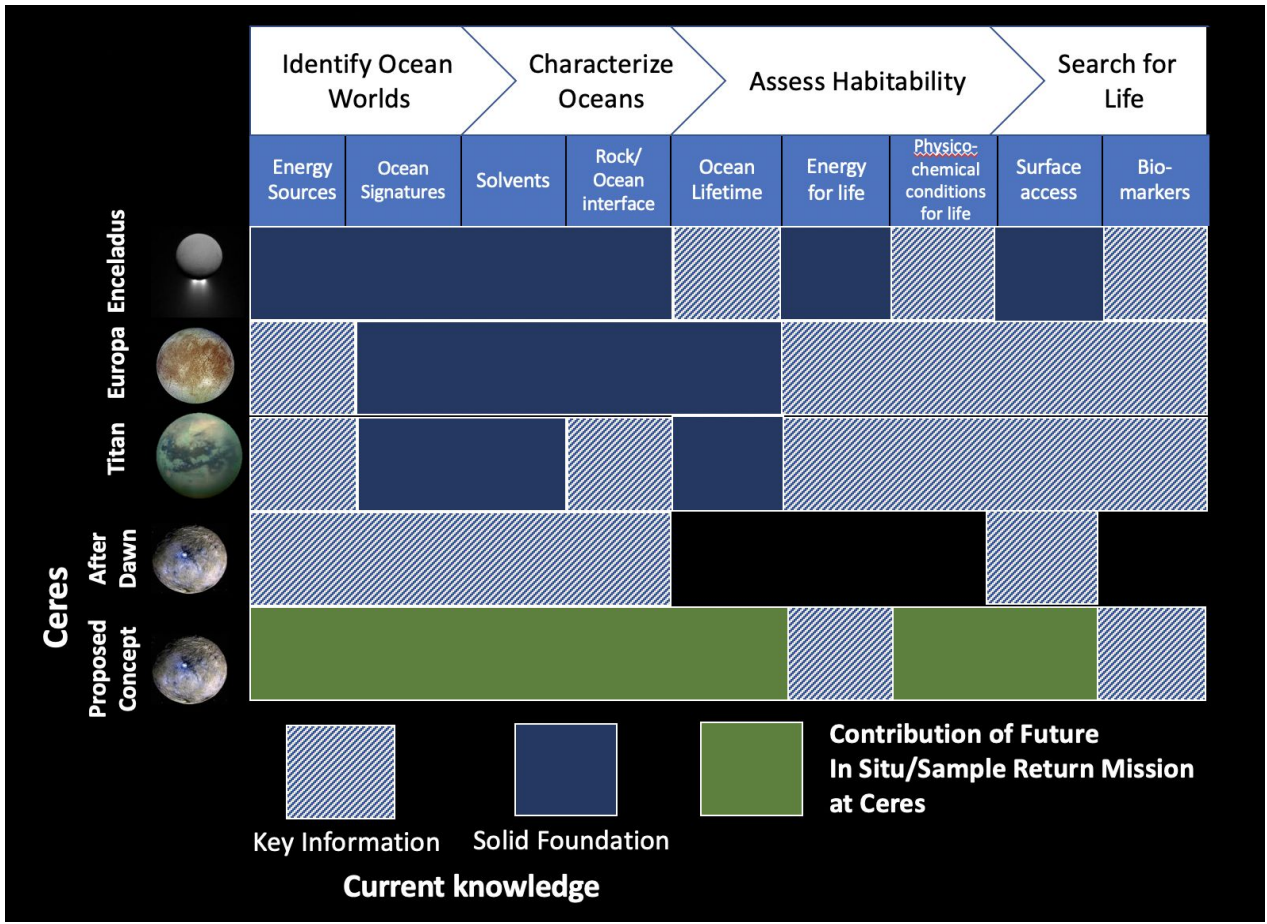

Figure 9. The contribution of future Ceres in-situ and sample return mission to the exploration of the solar system ocean worlds and their habitability.

Table 2. Traceability matrix for Project GAUSS

| Scientific goal | Measurement objective | Instrument |
|---|---|---|
| The origin and transportation of water and other volatiles in the inner solar system: Where does Ceres come from? | | |
| Connection between Ceres composition and pre-solar materials | Stable isotopes of oxygen, chromium, titanium | Sampling mechanism, microscopic camera |
| Volatile inventory on Ceres | Surface and subsurface water, other volatile species including Na, K, S, Cl, etc. | Sampling mechanism, NIR, TIR, APXS, GRS |
| Contamination from exogenous materials on Ceres and their roles in the evolution of Ceres | Surface mineralogy, petrological units in samples | Sampling mechanism, camera, NIR, TIR, microscopic camera |
| Physical properties and internal structure of Ceres: What do ice dwarf planets look like? | | |
| Structure of the near-subsurface and deep interior, and the implications to the differentiation and aqueous alteration processes of Ceres | Surface mineralogy, elemental abundance, gravity | Imaging, NIR, TIR, seismology, radar, subsurface science package |
| Geological processes, in particular current and past cryovolcanism and geothermal activity | Topography and morphology, gravity | Imaging, lidar, radar, radio science, seismometer |

| Ceres chronology | Crater counting, radiometric dating of samples | Sample analysis, camera, topographic camera |
|---|---|---|
| The existence and characteristics of Ceres exosphere | Gas species around Ceres | GC-MS, Ion and mass spectrometer, UV spectrometer, dust detector |
| The astrobiological implications of Ceres: Was it habitable in the past and is it still today? | | |
| Existence of liquid water inside Ceres, its extent, distribution, and depth | Topography, morphology, gravity | Camera, radar, lidar, topographic camera, radio science, seismometer, subsurface science package |
| Redox condition on Ceres, the existence and forms of oxidants | Elemental abundance, forms, and isotopic ratios of O, S, Cl, N | Sampling mechanism, NIR, TIR, APXS, GRS |
| Abundance, sources and sinks, and chemical forms of life-forming elements; source of terrestrial water | Elemental abundance, forms, and isotopic ratios of C, H, N, O, S | Sampling mechanism, NIR, TIR, APXS, GRS |
| Inventory and composition of organic compounds, their origins and evolutions | Existence, abundance, and types of organic materials | Sample mechanism, NIR, TIR |
| Mineralogical connection between Ceres and the collections of primitive meteorites: Where are Ceres meteorites? | | |
| Thermal metamorphism and aqueous alteration history of Ceres | Mineralogy, petrological characterization, isotope ratios | Sampling mechanism, NIR, TIR |
| Fractionation of elements and the geochemical processes | Mineralogy, elemental abundance, isotope ratios | Sampling mechanism, NIR, TIR |

## 3.2 Candidate sites for in-situ investigation and sample return

The geomorphological features on the surface of Ceres are connected to its interior and formed through recent geological processes (Fig. 10). The mantle of Ceres is composed of hydrated minerals, and may be enriched of soluble organics and high-density, low melting point organic matters in localized areas. The ~40-km thick crust above the mantle is composed of the original ocean materials, salts, carbonates, and brine. Liquid brine pockets existed in the recent past and drove cryovolcanic and geothermal activity. Ahuna Mons might be an extrusion feature of briny mud and organics from mantle. The surface layer of crust is covered by a mixture of infalls, salts, and organics. Cerealia Facula is covered by salts that are left behind after subsurface liquid brine was activated by impact and reached the surface through the cracks and eventually evaporated. Near young impact crater are freshly exposed materials that could be the original ocean materials.

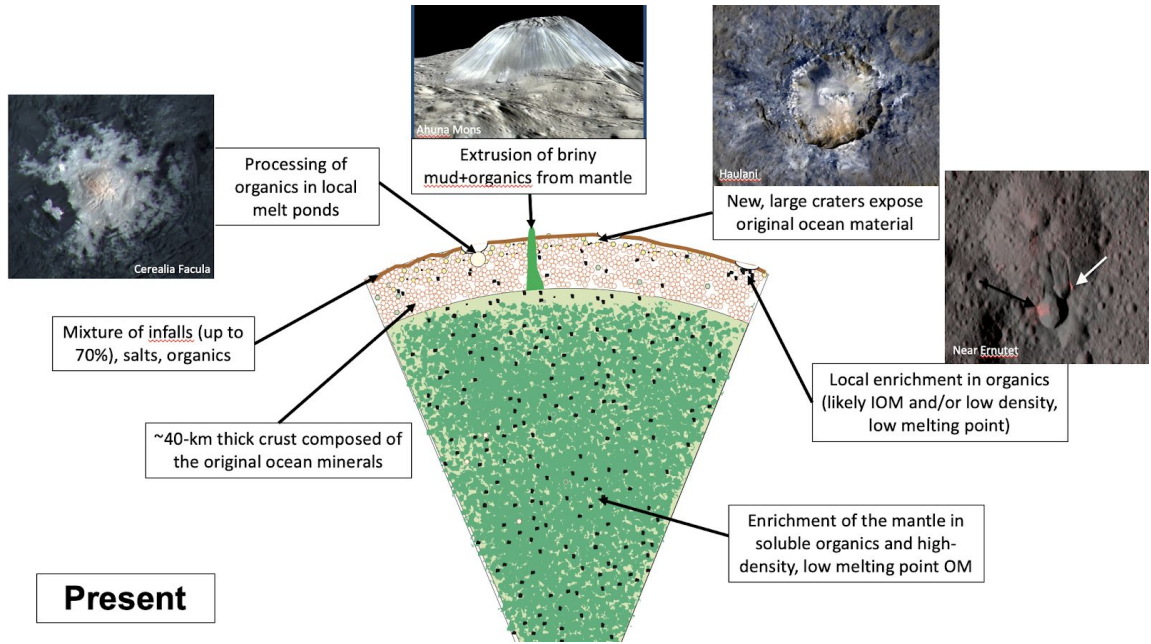

**Figure 10**. Surface geomorphological features on Ceres and their connections to the interior structure.

Therefore, Dawn observations of Ceres clearly pointed to those areas on Ceres as the potential sites that carry profound scientific implications serving to the scientific goals identified for Project GAUSS for in-situ investigations and sample returns (Fig. 11). The 92 km-diameter Occator crater is the host of Cerealia Facula, a dome with bright Na-rich carbonates deposits on the surface that is formed by recent geothermal activity. Ahuna Mons is the highest mountain and the only ascertained cryovolcanic feature that likely has a deep seated brine pocket beneath it as the driving force. Crater Ernutet is the most organic-rich area on Ceres. And Haulani is one of the youngest impact craters, associated with bright blue rays of ejecta that are freshly exposed crustal materials and could be associated with the original ocean materials.

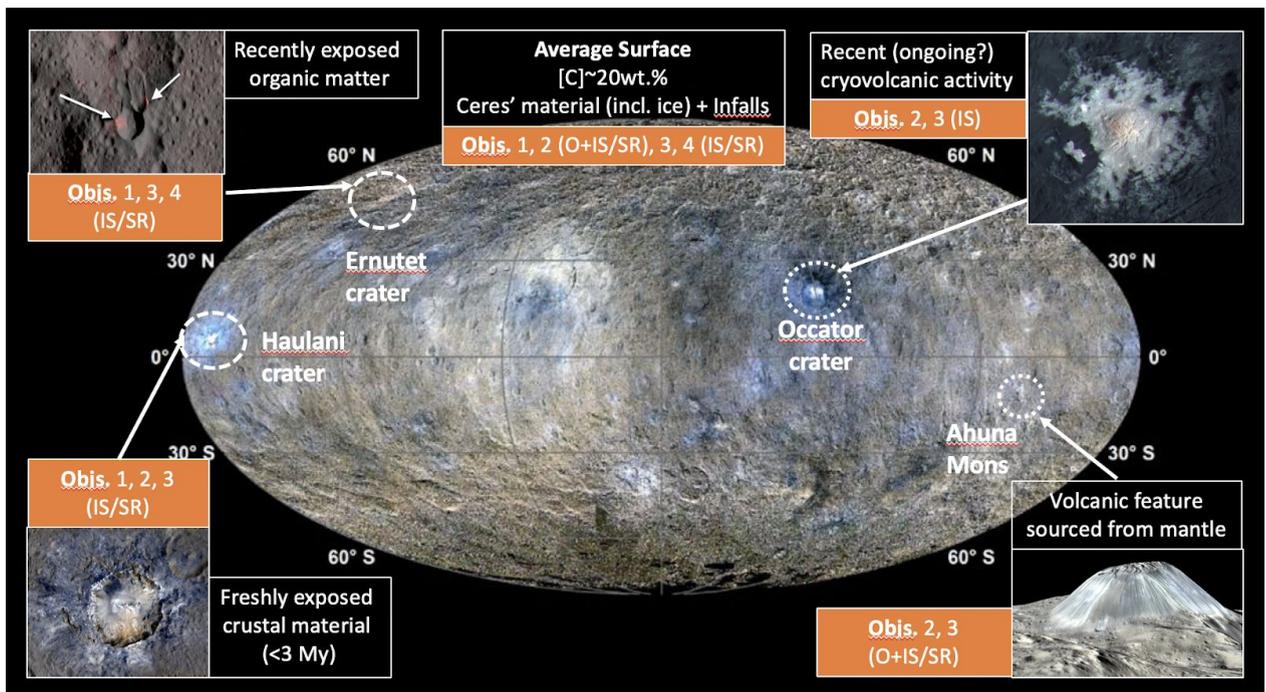

**Figure 11**. Candidate sites on Ceres for in-situ and/or sample return. The orange boxes list the scientific goals best served by these sites. O: orbiter; IS: in-situ measurements; SR: sample return. The Ceres basemap is a composite color map of Ceres generated by Dawn Framing Camera images with R, G, B = 960, 550, 440 nm, respectively.

## 3.3 Proposed payloads

Only three instruments were carried on Dawn. A more comprehensive payload would be needed to characterize Ceres itself and its atmospheric and space environment. This is especially true with the detection of water plume activity by Herschel (Küppers et al., 2014). Similar to the Cassini measurements at Enceladus, repeated fly-through of the gas plume would allow the identification of the chemical composition and isotopic ratios of the gas molecules, thus providing a probe to the nature of the subsurface lake/ocean.

The proposed payloads and their heritage are summarized in Tab. 3. An important assumption is made here. That is, it is assumed that some of the scientific instruments will have been successfully developed for CNSA's HX-1 mission to Mars and the China Asteroid Explorer (CAEX) to small bodies. By the same token, the sampling and reentry technology are also assumed to be derived from China's sampling missions to the Moon, Mars and asteroids. The instrumentation on the ESA's Philae lander of Rosetta, the Rosalind Franklin Rover (RFR) of ExoMars, the Yutu rovers of Chang'e missions, and the HX-1 Mars rover will provide strong preparation and heritage for the Ceres rover.

It is important to emphasize the astrobiological significance of the in-situ exploration of geological landmarks of special interest. These include the Ernutet crater and the Occator crater. To achieve high confidence in detecting organic materials on the surface of Ceres and assess their nature, the mid-IR range 2-12 µm turns out to be crucial at the spatial resolution that could be achieved by a spacecraft in orbit around Ceres, even with no imaging capabilities.
Covering this sensitivity range would in fact allow:

- Complementing surface mineralogy as derived by VIR in the overall 0.4-5.1 µm spectral range by using thermal emission spectroscopy from 2 to 12 µm, determining the specific compound responsible for the organics-rich area observed close to crater Ernutet, and shedding light on its origin (endogenous vs. exogenic). This spectral range would ultimately provide the capability to greatly expand our inventory of astrobiologically important compounds, and to remotely sense complex organics.
- Assess any potential ongoing activity and occurrence of organics, particularly on top of Cerealia Facula in crater Occator.
- Perform an in-depth characterization of the 34-km crater Haulani, one of the youngest geologic features on Ceres and home to the most prominent thermal signature observed on the entire surface of the dwarf planet (Tosi et al., 2018, 2019).
- Retrieve surface temperatures, grain size, porosity and surface roughness using thermal emission from 2 to 12 µm, thus accessing temperature values below 180 K with high accuracy, which were precluded to Dawn/VIR.
- Ultimately characterize the best landing site for a surface element in terms of composition and roughness (regolith depth).

Fourier Transform Spectrometer (FTS) working principle is based on the Michelson interferometer, which is an infrared spectrometer suitable for covering a broad spectral range from the near infrared to the mid infrared with constant, high spectral resolution (typically up to 1-2 cm$^{-1}$). FTSs simultaneously detect light varying the optical path difference (OPD) and encoding the signal at each wavelength with a cosine modulation at a frequency proportional to the wavenumber ν=1/λ. This interferogram is then transformed into a spectrum, i.e., from the optical path domain to the optical frequency domain. The OPD could be produced by rotating a double pendulum system around its axis, rather than translating the moving mirror along a linear direction like in the classical Michelson interferometer. Compared to the first generation of FTS (e.g. PFS on board the ESA Mars Express spacecraft), in recent years updated versions have been proposed that foresee a significant reduction in mass and size, adopting innovative technical solutions.

**Table 3.** Strawman payloads on the Orbiter and the Lander

|  | **Name** | **Heritage** |
|---|---|---|
| Orbiter | Wide-angle and narrow-angle camera | Chang'e 1-3, Dawn, Rosetta, Bepi Colombo |
|  | Infrared imaging spectrometer | Dawn, Rosetta, Chang'e-4, CAEX |
|  | Fourier Transform Spectrometer | Mars Express, ExoMars |
|  | Thermal mapper | CAEX |
|  | Ultraviolet imaging spectrometer | Chang'e-3, MMX |
|  | Gamma-ray spectrometer | Chang'e 1-2 |
|  | Long-wavelength radar | CAEX |
|  | Lidar | BepiColombo |
|  | Laser-induced breakdown spectroscopy | HX-1 |
|  | Dust detector | Rosetta, CAEX |
|  | Ion and mass spectrometer | Rosetta, CAEX |
|  | Particles & Fields package | Rosetta, CAEX |
|  | Radio Science | Rosetta |
| Lander | Topographic camera system | Chang'e 3-4, RFR |
|  | Microscopic camera | Philae, CAEX |
|  | Active Particle-induced X-ray Spectrometer (APXS) | Chang'e 1-3 |
|  | Gamma-ray spectrometer (GRS) | Chang'e 1-2 |
|  | Gas chromatography mass spectrometry (GC-MS) | Philae, CAEX |
|  | Subsurface science package | Philae, CAEX |
|  | Active seismometer | InSight, Chang'e-7 |

## 3.4 Trajectory design

Considering the fact that the target asteroid is located between the orbits of Mars and Jupiter (the semi-major axis of Ceres is about 2.768 au), in the transfer design we take advantage of a gravity assist of Mars in order to reduce the required fuel consumption. In the preliminary transfer scenario, the sequence is given as follows: (a) launching from the Earth; (b) Mars' gravity assist; (c) rendezvous with Ceres; (d) returning to the Earth. The launch window is assumed to be after January 1st, 2035. The planar transfer trajectory is reported in Fig. 12 and the associated parameters are provided in Tab. 4.

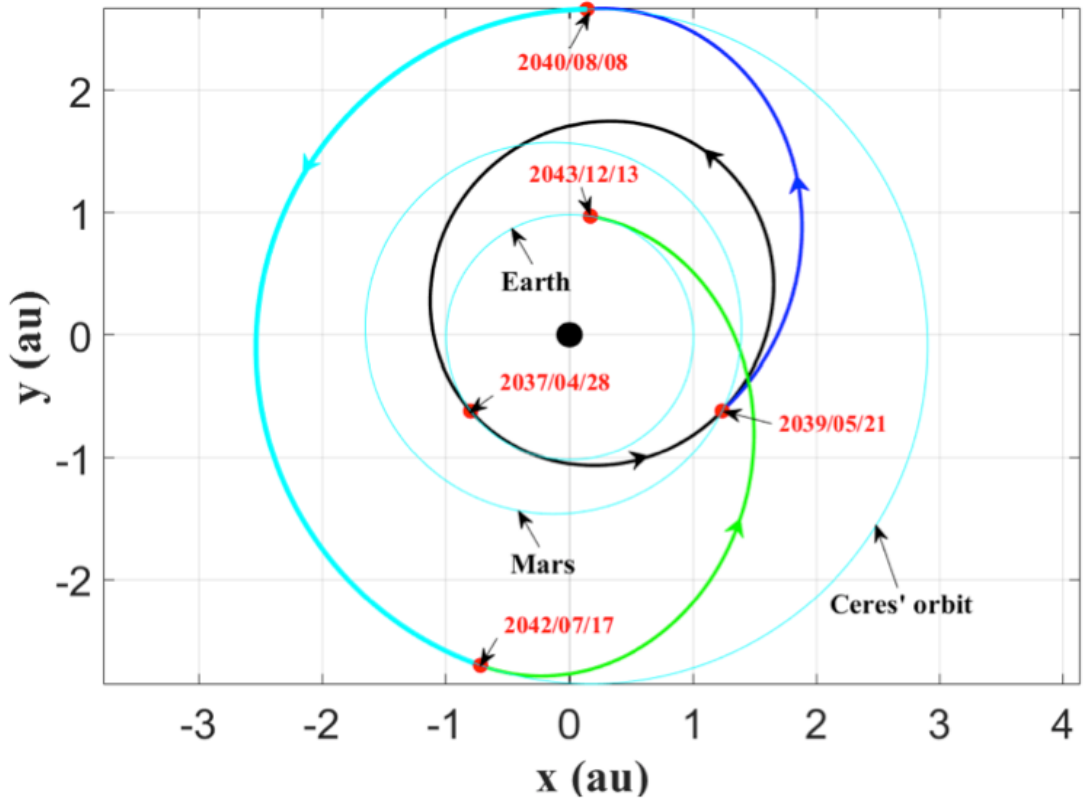

**Figure 12.** The transfer trajectory for Ceres' sample return mission. In the transfer scenario, a Mars' gravity assist is taken into consideration. The probe launches from the Earth on April 28th, 2037. The arc from the Earth to Mars is shown in black line, the arc from Mars to the target asteroid is given in blue line, the rendezvous segment is presented in cyan line, and the return trajectory from the asteroid to the Earth is marked in green line. The critical points of time are represented by red dots.

**Table 4.** Parameters of the interplanetary trajectory for the Ceres sample return mission

| Time points | | Velocity impulse | |
|---|---|---|---|
| Launch | April 28, 2037 | Hyperbolic excess velocity ($v_\infty$) | 4.961 km/s |
| Mars' gravity assist | May 21, 2039 | Deep space maneuver at perimartian ($\Delta v_M$) | 1.981 km/s |
| Rendezvous with Ceres | August 8, 2040 | Braking velocity for rendezvous with Ceres ($\Delta v_f$) | 5.400 km/s |
| Departure from Ceres | July 17, 2042 | Accelerating velocity for departing from Ceres | 4.813 km/s |
| Earth re-entry | December 13, 2043 | | |

## 3.5 Technological requirements

Here we discuss the key technological requirements and main challenges for a Ceres sample return mission:

- **Flight dynamics**
  It is assumed, that by the time of program "Voyage 2050", the capacity of launch vehicles poses no challenge for sending a spacecraft with a lander and re-launch system to Ceres. Ceres does not have an atmosphere that exhibits any detectable dynamical effects on Dawn spacecraft at altitudes down to 35 km. Therefore entry into the atmosphere necessary for Mars landing is not needed to land on Ceres. The descent and landing system for Ceres can be derived from the already mature Chinese Chang'E 3 and 4 descent and landing system. The technology for re-launch system already exists since the Apollo era, and a similar system is already in development for the various Mars sample return missions in study (see *Visions and Voyages*, the Planetary Science Decadal Survey 2013-2022 report, National Research Council of the National Academies). The escape velocity from Ceres' surface of about 0.5 km/s is much lower than that of the Moon (2.4 km/s) and Mars (5 km/s). Therefore, the re-launch system from Ceres' surface for sample return can be much smaller and more light-weight than those for the Apollo system (which also had a complicated life-support system) or any Mars sample return system in development. It can be expected that the needed technology will have been developed and matured by 2030-2040 timeframe, and should be much more efficient (small and light-weight) than the currently available system that will require a less powerful launch vehicle than required by the present technology. Earth reentry is already mature in today's technology and is not expected to pose any challenge. In addition, an ion propulsion system, like the one used in the Dawn mission but should be much more efficient by then, might also be adopted to boost the velocity increment.

- **Sampling on Ceres**
  One or more approaches need to be developed for sampling the surface layer of Ceres. Sampling on Ceres is different from that on the Moon or Mars, which have relatively strong surface gravity and an anchoring mechanism may not be necessary. Ceres sampling would be different from that on small asteroids such as the targets of JAXA's Hayabusa and Hayabusa2, and NASA's OSIRIS-REx, which have loose regolith and micro-gravity allowing for a touch-and-go approach. With its surface gravitational acceleration of ~0.27 m/s$^2$, a robust anchoring mechanism is required for sampling on Ceres, and a drilling system is also needed to drill into the relatively strong rock-ice-salt mixture with a density of ~1.7 g/cm$^3$ (Park et al. 2016) to collect ice-rich samples from subsurface. A drilling depth of decimetres, but less than one metre is likely needed in order to reach the water-ice rich layers or subsurface ice table (Prettyman et al. 2017). Technologies currently being developed for landing on and sampling asteroids might provide the necessary technological readiness for project GAUSS, e.g., technologies being developed for the Chinese small body mission (Zhang et al. 2017). The various sampling mechanisms and collection system for Mars sample return currently in development (e.g., Magnani et al. 2011, Zacny 2014) are good references for the future development of Ceres sample return system. Note that ESA and NASA signed a statement of intent on Mars sample return in April 2018.

- **Cryogenic sample collection, containment, and curation**
  This is probably the most challenging part of the entire project. To preserve the volatile and organics in their original status, samples should not be thermally or aqueously altered during the collection, storage and transport, and if possible, not even mechanically altered. Therefore, the samples need to be collected and sealed in the containers with cryogenic temperature and overpressure. A temperature of <~170 K is required, and the original temperature of subsurface samples (probably ~140 K, Schorghofer 2008) is desired, throughout the entire collection, return, and curation process in order to prevent water ice from melting and aqueously alter any mineral samples. The sample return capsule should have a number of separated and individually sealed containers that can be overpressured to prevent the loss of volatiles such as H2O, HCN, S-bearing species, and cyanides (Chuang et al. 2019, Chiang et al. 2019). It is worth mentioning that Rosetta was originally designed to return cryogenic samples from a comet (Huber & Schwehm 1991).

# 4. International context

The extensive orbital measurements of the Dawn mission can be utilized to the full advantage in planning this large-scale mission of unique importance. Because of the efforts of several agencies including JAXA, NASA, and ESA over the last two decades, asteroidal exploration has become a major component of deep-space missions. On the one hand, sample return missions and the related technology have been championed by JAXA that has successfully executed the Hayabusa mission to the S-type asteroid Itokawa and the Hayabusa 2 mission to a C-type asteroid Ryugu. This line of approach will be extended to the sample return mission (Mars Moon Explorer or MMX) to the Martian moon Phobos and very possibly a sample return mission to a Jovian Trojan asteroid. Following the Dawn mission, NASA has demonstrated its strong interest in asteroid in-situ exploration by running the OSIRIS-ReX sample-collection project to the B-type asteroid Bennu, and organizing two more asteroid rendezvous missions, one to the large M-type (iron) asteroid, Psyche, and the other one (Project Lucy) to a number of Jovian Trojans. The Chinese space agency, CNSA, has recently issued an AO calling for instrument proposals and international cooperation for its first asteroid sample return mission to be followed by a rendezvous mission to a main-belt asteroid that exhibited outgassing activity. There is also an assessment study for a sample return mission to an inner-belt E-type asteroid. In parallel to this heightened level of activities by JAXA, NASA and CNSA, ESA has approved the daring "Comet Interceptor" project at the heel of the Rosetta/Philae mission.

If the above set of planned (or proposed) space projects can be successfully carried out (or launched), the first phase of reconnaissance/sample-return missions to major phenotypes of small asteroids and short-period comets - as a global enterprise - would be completed by 2030-2035. It is also expected that the basic knowledge gained and technologies invented will be applied to the next phase of asteroidal exploration - beginning around 2030 if not earlier - with a view to address the need of Earth defense against asteroid impact hazard - and asteroidal mining for commercial reasons.

Extended proposing team

John Carter (Institut d'Astrophysique Spatiale), Grégoire Danger (Aix Marseille Univ, CNRS), Julia de Leon (Instituto de Astrofisica de Canarias), Jörn Helbert (DLR Institute of Planetary Research), Xiyun Hou (Nanjing University), Hauke Hussmann (DLR Institute of Planetary Research), Katherine Joy (University of Manchester), Tomas Kohout (University of Helsinki), Alice Lucchetti (INAF-Osservatorio astronomico di Padova), David Mimoun (Université de Toulouse), Olga Muñoz (Instituto de Astrofisica de Andalucia), Jose Luis Ortiz (Instituto de Astrofisica de Andalucia), Antti Penttilä (University of Helsinki), Frank Preusker (DLR Institute of Planetary Research), Ottaviano Ruesch (University of Münster), Pablo Santos-Sanz (Instituto de Astrofisica de Andalucia), Nico Schmedmann (University of Münster), Nicole Schmitz (DLR Institute of Planetary Research), Katrin Stephan (DLR Institute of Planetary Research), Guneshwar Thangjam (NISER), Josep Maria Trigo-Rodriguez (CSIC-IEEC), Cecilia Tubiana (Max Planck Institute for Solar System Research), Vassilissa Vinogradoff (Aix Marseille University, CNRS), Liangliang Yu (Macau University of Science and Technology), Francesca Zambon (INAF-IAPS), Yuhui Zhao (Purple Mountain Observatory, CAS)

**[References]**

A'Hearn, M.F., Feldman, P.D., 1992. Water vaporization on Ceres. Icarus 98, 54-60.

Altwegg, K., Balsiger, H., Bar-Nun, A., et al. 2015. 67P/Churyumov-Gerasimenko, a Jupiter family comet with a high D/H ratio. Science, 347, 1261952

Ammannito, E., et al. 2016. Distribution of phyllosilicates on the surface of Ceres. Science 353, aaf4279.

Bland, M.T., et al. 2016. Composition and structure of the shallow subsurface of Ceres revealed by crater morphology. Nature Geosci. 9, 538-542.

Bottke, W. F., Vokrouhlicky, D., Rubincam, D. P., Broz, M. 2002. The effect of Yarkovsky thermal forces on the dynamical evolution of asteroids and meteoroids. In Asteroids III, 395-408.

Bottke, W. F., Durda, D. D., Nesvorny, D., et al. 2005. Linking the collisional history of the main asteroid belt to its dynamical excitation and depletion. Icarus, 179, 63-94.

Bottke, W. F., Vokrouhlicky, D., Rubincam, D. P., Nesvorny, D. 2006. The Yarkovsky and YORP effects: implications for asteroid dynamics. Annu. Rev. Earth Planet. Sci., 34, 157–191.

Buczkowski, D.L., et al. 2016. The geology of Ceres. Science 353, id.aaf4332.

Buczkowski, D.L., et al. 2019. Tectonics analysis of fracturing associated with Occator crater. Icarus 320, 49-59.

Budde, G., Burkhardt, C., Brennecka, G.A., Fischer-Gödde, M., Kruijer, T.S. and Kleine, T., 2016. Molybdenum isotopic evidence for the origin of chondrules and a distinct genetic heritage of carbonaceous and non-carbonaceous meteorites. *Earth and Planetary Science Letters*, *454*, pp.293-303.

Campins, H., Hargrove, K., Pinilla-Alonso, N., et al. 2010. Water ice and organics on the surface of the asteroid 24 Themis. Nature, 464, 1320-1321.

Castillo-Rogez J., et al. (2018) Insights into Ceres' evolution from surface composition. Meteorit. Planet. Sci. 53, 1820-1843, doi:10.1111/map.13181.

Chiang, C.-C., Kuan, Y.-J., Chuang, Y.-L., 2019. ACA observations of Ceres' molecular exosphere. 16th AOGS Meeting, Singapore, abstract #PS16-A023.

Chuang, Y.-L., 2019. Submillimeter spectral observations of molecular exosphere of Ceres icy world. 16th AOGS Meeting, Singapore, abstract #PS16-A025.

Ciesla, F. J., Cuzzi, J. N. 2006. The evolution of the water distribution in a viscous protoplanetary disk. Icarus, 181, 178-204.

Combe, J.-P., et al. 2016. Detection of local H2O exposed at the surface of Ceres. Science 353, id.aaf3010.

Combe, J.-P., et al. 2019. Exposed H2O-rich areas detected on Ceres with the Dawn visible and infrared mapping spectrometer. Icarus 318, 22-41.

Connelly, J.N., Amelin, Y., Krot, A.N. and Bizzarro, M., 2008. Chronology of the solar system's oldest solids. The Astrophysical Journal Letters, 675(2), p.L121.

De Sanctis, M.C., et al. 2015. Ammoniated phyllosilicates with a likely outer solar system origin on (1) Ceres. Nature 528, 241-244.

De Sanctis M.C., et al. 2016. Bright carbonate deposits as evidence of aqueous alteration on (1) Ceres. Nature 536, 54-57.

De Sanctis, M.C., et al. 2017. Localized aliphatic organic material on the surface of Ceres. Science 355, 719-722.

DeMeo, F. E., Carry, B. 2013. The taxonomic distribution of asteroids from multi-filter all-sky photometric surveys. Icarus, 226, 723-741.

DeMeo, F. E., Carry, B. 2014. Solar System evolution from compositional mapping of the asteroid belt. Nature 505, 629-634.

Ermakov, A.I., et al. 2017. Ceres' obliquity history and its implications for the permanently shadowed regions. Geophys. Res. Lett. 44, 2652-2661.

Farinella, P., Vokrouhlicky, D., Hartmann, W. K. 1998. Meteorite delivery via Yarkovsky orbital drift. Icarus, 132, 378–387.

Fernandez, J.A. and Ip, W.H., 1984. Some dynamical aspects of the accretion of Uranus and Neptune: The exchange of orbital angular momentum with planetesimals. Icarus, 58(1), pp.109-120.

Fu, R.R., et al. 2017. The interior structure of Ceres as revealed by surface topography. Earth Planet. Sci. Lett. 476, 153-164.

Gladman, B. J. et al. 1997. Dynamical lifetimes of objects injected into asteroid belt resonances. Science, 277, 197–201.

Gomes, R., Levison, H. F., Tsiganis, K., Morbidelli, A. 2005. Origin of the cataclysmic Late Heavy Bombardment period of the terrestrial planets. Nature, 435, 466-469.

Gradie, J., Tedesco, E. 1982. Compositional structure of the asteroid belt. Science, 216, 1405-1407.

Hahn, J. M., Malhotra, R. 1999. Orbital Evolution of Planets Embedded in a Planetesimal Disk. Astron. J., 117, 3041-3053.

Hargrove, K. D., Emery, J. P., Campins, H., et al. 2015. Asteroid (90) Antiope: Another icy member of the Themis family? Icarus, 254, 150-156.

Hartogh, P., et al. 2011. Ocean-like water in the Jupiter-family comet 103P/Hartley 2. Nature, 478, 218-220

Hendrix, A.R., et al., 2019. The NASA roadmap to ocean worlds. Astrobiology 19, 1-27.

Hiesinger, H., et al. 2016. Cratering on Ceres: Implications for its crust and evolution. Science 353, id.aaf4758.

Hiroi, T., et al. 1996, "Thermal metamorphism of the C, G, B, and F asteroids in comparison with carbonaceous chondrites", Meteoritics & Planetary Science, 31, 321-327.

Hsieh, H. H., Jewitt, D. 2006, "A population of comets in the main asteroid belt", Science, 312, 561-563

Keil, K. 2000, "Thermal alteration of asteroids: evidence from meteorites", Planet. & Space Sci., 48, 887-903

Huber, M.C. and Schwehm, G., 1991. Comet nucleus sample return: plans and capabilities. Space Science Reviews, 56(1-2), pp.109-115.

Jacobson, S.A., Morbidelli, A., Raymond, S.N., O'Brien, D.P., Walsh, K.J. and Rubie, D.C., 2014. Highly siderophile elements in Earth's mantle as a clock for the Moon-forming impact. Nature, 508(7494), p.84.

Jones, T. D., Lebofsky, L. A., Lewis, J. S., Marley, M. S. 1990. The composition and origin of the C, P, and D asteroids: Water as a tracer of thermal evolution in the outer belt. Icarus, 88, 172-192.

King. S.D. et al. 2018. Ceres internal structure from geophysical constraints. Meteorit. Planet. Sci. 53, 1999-2007.

Krohn, K. et al. 2016. Cryogenic flow features on Ceres: Implications for crater-related cryovolcanism. Geophys. Res. Lett. 43, 11,994-12,003.

Kruijer, T.S., Burkhardt, C., Budde, G. and Kleine, T., 2017. Age of Jupiter inferred from the distinct genetics and formation times of meteorites. *Proceedings of the National Academy of Sciences*, *114*(26), pp.6712-6716.

Küppers, M., et al. 2014. Localized sources of water vapour on the dwarf planet (1) Ceres. Nature 505, 525-527.

Landis, M.E., et al. 2017. Conditions for sublimating water ice to supply Ceres' exosphere. J. Geophys. Res. 122, 1984-1995.

Landis, M.E., et al. 2019. Water vapor contribution to Ceres' exosphere from observed surface ice and postulated ice-exposing impacts. J. Geophys. Res. Planet. 124, 61-75.

Lecar, M., Podolak, M., Sasselov, D., Chiang, E. 2006. On the location of the snow line in a protoplanetary disk. Astrophys. J., 640, 1115-1118.

Levison, H. F., Bottke, W. F., Gounelle, M., Morbidelli, A., Nesvorny, D., Tsiganis, K. 2009. Contamination of the asteroid belt by primordial trans-Neptunian objects. Nature, 460, 364-366.

Levison, H. F., Morbidelli, A., Tsiganis, K., Nesvorny, D. Gomes, R. 2011. Late orbital instabilities in the outer planets induced by interaction with self-gravitating planetesimal disk. Astron. J. 142, 152.

Magnani, P., et al., 2011. Testing of ExoMars EM drill tool in Mars analogous materials. In : Proceedings of ESA/ESTEC, pp. 1-8.

Marchi, S., Barbieri, C., Küppers, M., Marzari, F., Davidsson, B., Keller, H.U., Besse, S., Lamy, P., Mottola, S., Massironi, M. and Cremonese, G., 2010. The cratering history of asteroid (2867) Steins. Planetary and Space Science, 58(9), pp.1116-1123.

Marion, G.M., Kargel, J.S., Catling, D.C., Lunine, J.I., 2012. Modeling ammonia-ammonium aqueous chemistries in the solar system's icy bodies. Icarus 220, 932-946.

McKinnon, W.B., 2012. Where did Ceres accrete – In situ in the asteroid belt, among the giant planets, or in the primordial transneptunian belt? AAS DPS meeting #44, id.111.14.

McKinnon, W.B., Zolensky, M.E., 2003. Sulfate content of Europa's ocean and shell: Evolutionary considerations and some geological and astrobiological implications. Astrobiology 3, 879-897.

McSween H. Y., et al. (2018) Carbonaceous chondrites as analogs for the composition and alteration of Ceres. Meteorit. Planet. Sci. 53, 1793-1804, doi:10.1111/maps.13124.

Morbidelli, A., Levison, H. F., Tsiganis, K. & Gomes, R. 2005. Chaotic capture of Jupiter's Trojan asteroids in the early Solar System. Nature, 435, 462-465.

Nanne, J.A., Nimmo, F., Cuzzi, J.N. and Kleine, T., 2019. Origin of the non-carbonaceous–carbonaceous meteorite dichotomy. *Earth and Planetary Science Letters*, *511*, pp.44-54.

Nathues, A., et al., 2017. Evolution of Occator crater on (1) Ceres. Astron. J. 153, 112 (12pp).

Nathues, A., et al., 2019. Occator crater in color at highest spatial resolution. Icarus 320, 24-38.

Neesemann, A., et al., 2019. The various ages of Occator crater, Ceres: Results of a comprehensive synthesis approach. Icarus 320, 60-82.

Neveu, M. and Desch, S.J., 2015. Geochemistry, thermal evolution, and cryovolcanism on Ceres with a muddy ice mantle. Geophysical Research Letters, 42(23), pp.10-197.

O'Brien, D. P., Izidoro, A., Jacobson, et al. 2018, "The Delivery of Water During Terrestrial Planet Formation", Space Sci. Rev., 214, 47

Park, R.S., Konopliv, A.S., Bills, B.G., Rambaux, N., Castillo-Rogez, J.C., Raymond, C.A., Vaughan, A.T., Ermakov, A.I., Zuber, M.T., Fu, R.R. and Toplis, M.J., 2016. A partially differentiated interior for (1) Ceres deduced from its gravity field and shape. Nature, 537(7621), p.515.

Poole, G.M., Rehkämper, M., Coles, B.J., Goldberg, T. and Smith, C.L., 2017. Nucleosynthetic molybdenum isotope anomalies in iron meteorites–new evidence for thermal processing of solar nebula material. *Earth and Planetary Science Letters*, *473*, pp.215-226.

Postberg, F., Schmidt, J., Hillier, J., Kempf, S., Srama, R., 2011. A salt-water reservoir as the source of a compositionally stratified plume on Enceladus. Nature 474, 620-622.

Prettyman,T.H. et al., 2017. Extensive water ice within Ceres' aqueously altered regolith: Evidence from nuclear spectroscopy. Science 355, 55-59.

Qin, L., Nittler, L.R., Alexander, C.O.D., Wang, J., Stadermann, F.J. and Carlson, R.W., 2011. Extreme 54Cr-rich nano-oxides in the CI chondrite Orgueil–Implication for a late supernova injection into the solar system. Geochimica et Cosmochimica Acta, 75(2), pp.629-644.

Quick, L.C., et al., 2019. A possible brine reservoir beneath Occator crater: Thermal and compositional evolution and formation of the Cerealia dome and Vinalia Faculae. Icarus 320, 119-135.

Raponi, A., et al., 2019. Mineralogy of Occator crater on Ceres and insight into its evolution from the properties of carbonates, phyllosilicates, and chlorides. Icarus 320, 83-96.

Rivkin, A. S., Howell, E. S., Vilas, F., Lebofsky, L. A. 2002. Hydrated Minerals on Asteroids: The Astronomical Record. In Asteroids III 235.

Rivkin, A. S., & Emery, J. P. 2010, "Detection of ice and organics on an asteroidal surface", Nature, 464, 1322-1323

Rivkin, A.S., Thomas, C.A., Trilling, D.E., Enga, M.T. and Grier, J.A., 2011. Ordinary chondrite-like colors in small Koronis family members. Icarus, 211(2), pp.1294-1297.

Ruesch, O., et al., 2016. Cryovolcanism on Ceres. Science 353, id.aaf4286.

Ruesch, O., et al., 2019. Slurry extrusion on Ceres from a convective mud-bearing mantle. Nature Geosci. 12, 505-509.

Russell C.T., et al., 2016. Dawn arrives at Ceres: Exploration of a small, volatile-rich world. Science 353, 1008-1010.

Sasselov, D. D., Lecar, M. 2000. On the snow line in dusty protoplanetary disks. Astrophys. J., 528, 995-998.

Schmidt, B.E., et al., 2017. Geomorphological evidence for ground ice on dwarf planet Ceres. Nature Geosci. 10, 338-343.

Schorghofer, N., 2008. The lifetime of ice on main belt asteroids. Astrophys. J. 682, 697-705.

Schorghofer, N., 2016. Predictions of depth-to-ice on asteroids based on an asynchronous model of temperature, impact stirring, and ice loss. Icarus 276, 88-95.

Schorghofer, N., et al., 2016. The permanently shadowed regions of dwarf planet Ceres. Geophys. Res. Lett. 43, 6783-6789.

Schorghofer, N., et al., 2017. The putative cerean exosphere. Astrophys. J. 850, 85 (7pp).

Scott, E. R. D., Taylor, G. J., Newsom, H. E., Herbert, F. L., Zolensky, M., Kerridge, J. F. 1989. Chemical, thermal and impact processing of asteroids. In Asteroids II, 701-739.

Scott E. R. D., et al. (2018) Isotopic dichotomy among meteorites and its bearing on the protoplanetary disk. Astrophys. Jour. 854, 164-176.

Scully, J.E., et al., 2019a. Ceres' Occator crater and its faculae explored through geologic mapping. Icarus 320, 7-23.

Scully, J.E., et al., 2019b. Synthesis of the special issue: The formation and evolution of Ceres' Occator crater. Icarus 320, 213-225.

Sizemore, H.G., et al., 2017. Pitted terrains on (1) Ceres and implications for shallow subsurface volatile distribution. Geophys. Res. Lett. 44, 6570-6578.

Sori, M.M., et al., 2017. The vanishing cryovolcanism of Ceres. Geophys. Res. Lett. 44, 1243-1250.

Sori, M.M., et al., 2018. Cryovolcanic rate on Ceres revealed by topography. Nature Astron. 2, 946-950.

Takir, D., Emery, J. P. 2012, "Outer main belt asteroids: Identification and distribution of four 3-μm spectral groups", Icarus, 219, 641-654

Teets, D. and Whitehead, K., 1999. The discovery of Ceres: How Gauss became famous. Mathematics Magazine, 72(2), pp.83-93.

Thomas, E.C., et al., 2019. Kinetic effect on the freezing of ammonium-sodium-carbonate-chloride brines and implications for the origin of Ceres' bright spots. Icarus 320, 150-158.

Tosi, F., et al., 2018. Mineralogy and temperature of crater Haulani on Ceres. Meteor. Planet. Sci. 53 (9), 1902-192.

Tosi, F., et al., 2019. Mineralogical analysis of the Ac-H-6 Haulani quadrangle of the dwarf planet Ceres. Icarus 318, 170-187.

Travis, B.J., Bland, P.A., Feldman, W.C., Sykes, M.V., 2018. Hydrothermal dynamics in a CM-based model of Ceres. Meteorit. Planet. Sci. 53, 2008-2032.

Tsiganis, K., Gomes, R., Morbidelli, A., Levison, H. F. 2005. Origin of the orbital architecture of the giant planets of the Solar System. Nature, 435, 459-461.

Villarreal, M.N., et al., 2017. The dependence of the cerean exosphere on solar energetic particle events. Astrophys. J. Lett. 838, 8 (5pp).

Walsh, K. J., Morbidelli, A., Raymond, S. N., O'Brien, D. P., Mandell, A. M. 2011. A low mass for Mars from Jupiter's early gas-driven migration. Nature, 475, 206-209.

Walsh, K.J., Morbidelli, A., Raymond, S.N., O'brien, D.P. and Mandell, A.M., 2012. Populating the asteroid belt from two parent source regions due to the migration of giant planets—"The Grand Tack". *Meteoritics & Planetary Science*, *47*(12), pp.1941-1947.

Warren, P.H., 2011. Stable-isotopic anomalies and the accretionary assemblage of the Earth and Mars: A subordinate role for carbonaceous chondrites. *Earth and Planetary Science Letters*, *311*(1-2), pp.93-100.

Weissman, P. R. 1999. Diversity of Comets: Formation Zones and Dynamical Paths. Space Science Reviews, 90, 301-311.

Williams D. A., et al. (2018) High-resolution global geologic map of Ceres from NASA Dawn mission. In Planetary Geologic Mappers Annual Meeting, Lunar and Planetary Institute, abstr. 7001.

Zacny, K., 2014. Drilling and caching architecture for the Mars2020 mission. In : Proceedings of the AGU Fall Meeting Abstracts, pp. 1797.

Zambon, F., et al., 2017. Spectral analysis of Ahuna Mons from Dawn mission's visible-infrared spectrometer. Geophys. Res. Lett. 44 (1), 97-104.

Zhang, T., Zhang, W., Wang, K., Gao, S., Hou, L., Ji, J. and Ding, X., 2017. Drilling, sampling, and sample-handling system for China's asteroid exploration mission. Acta Astronautica, 137, pp.192-204.